\documentclass[useAMS,usenatbib]{mn2e}
\usepackage{graphicx,rotate,url,mathptmx,lscape,times}
\usepackage{aas_macros}

\usepackage{color}
\usepackage{amsmath}
\usepackage{amssymb}
\usepackage{multirow}
\font\manual=manfnt at 7pt \def\dbend{\hbox{\raise0.9ex\hbox{\manual\char127\hspace{0.6em}}}}

\providecommand{\e}[1]{\ensuremath{\times 10^{#1}}}

\newcounter{INTERNALionstage}

\def\gtsim{\mathrel{\hbox{\rlap{\hbox{\lower4pt\hbox{$\sim$}}}\hbox{$>$}}}}
\def\lesssim{\mathrel{\hbox{\rlap{\hbox{\lower4pt\hbox{$\sim$}}}\hbox{$<$}}}}

\def\ga{{\rm\thinspace gauss}}

\def\la{\mbox{{\rm L}$\alpha$}}

\DeclareMathAlphabet{\vib}{OML}{cmm}{m}{it}
\newcommand{\Ubar}{\overline{U}}
\newcommand{\rme}{{\rm e}}
\newcommand{\kb}{k_{\rm B}}

\title[H, He-like recombination spectra II]{H, He-like recombination spectra II:  
$l$-changing collisions for He Rydberg states}

\author[F. Guzm\'an et al.]
       {\parbox[]{6.0in}
        { F. Guzm\'an$^1$, N. R. Badnell$^2$,  R. J. R. Williams$^{3}$, P. A. M. van Hoof$^4$, 
        M. Chatzikos$^1$ and G. J. Ferland$^{1}$. \\
        \footnotesize
        $^1$Department of Physics and Astronomy, University of Kentucky, Lexington, KY 40506, USA\\
        $^2$Department of Physics, University of Strathclyde, Glasgow G4 0NG, UK\\
        $^3$AWE plc, Aldermaston, Reading RG7 4PR, UK\\
        $^4$Royal Observatory of Belgium, Ringlaan 3, 1180 Brussels, Belgium}
}

\date{%Accepted .
      Received }

\pagerange{\pageref{firstpage}--\pageref{lastpage}}
\pubyear{}

\begin{document}

\maketitle

\label{firstpage}

\begin{abstract}

Cosmological models can be constrained by determining primordial abundances. Accurate 
predictions of the He~I spectrum are needed to determine the primordial helium 
abundance to a precision of $< 1$\% in order to constrain Big Bang Nucleosynthesis models.
Theoretical line emissivities at least this accurate are needed if this precision is to be achieved.
In the first paper of this series, which  focused on H~I, we showed that differences in $l$-changing
collisional rate coefficients predicted by three different theories can translate into 10\% changes
in predictions for H~I spectra. Here we consider the more 
complicated case of He atoms, where low-$l$ subshells are not energy degenerate.
A criterion for deciding when the energy separation between $l$ subshells is small
 enough to apply energy-degenerate collisional theories is given.
Moreover, for certain conditions, the Bethe approximation originally proposed by 
\cite{PengellySeaton1964} is not sufficiently accurate.  We introduce a simple modification
of this theory which leads to rate coefficients
which agree well with those obtained from pure quantal calculations using the approach of 
\cite{VOS2012}.
We show that the $l$-changing rate coefficients from the different theoretical approaches lead to 
differences of  $\sim 10$\% in He~I emissivities in simulations of H~II regions using spectral code 
Cloudy.
\end{abstract}

\begin{keywords}
atomic data -- ISM: abundances -- (ISM:) HII regions --
cosmology: observations -- (cosmology:) primordial nucleosynthesis

\end{keywords}

\section{Introduction}
\label{sec:intro}
Measurement of the primordial helium abundance  relies on the observation of H~I and He~I 
recombination lines in low-metallicity extragalactic H~II regions 
\citep{Izotov2007}.
The statistical uncertainties in derived He  abundances must be less than 1\% 
\citep{Olive2000} if they are to test Standard Big Bang 
Nucleosynthesis (SBBN) models \citep{Izotov2014}. 
This high precision must be 
supported by the fundamental atomic data employed in the H~I and He~I emissivity predictions. 

In the first paper of this series \citep{Guzman2016a}, hereafter P1, we examined 
how dipole $l$-changing collisions influence the H~I  spectrum. 
New semi-classical (SC) rate coefficients from \cite{VOS2012} (hereafter VOS12) 
are about an order of magnitude smaller  than those obtained using the 
long-standard theory of \cite{PengellySeaton1964} (hereafter PS64). 
The quantum mechanical (QM) treatment by \cite{Vrinceanu2001} is
in good agreement with PS64. 
The dipole collision probabilities of the PS64 and QM theories are divergent,
with the  cross section tending to infinity for collisions with very distant particles.
This necessitates the imposition of a cut-off at large impact parameters,
which in turn creates an additional uncertainty. 
The standard approach is to consider physical criteria such as collective effects (Debye screening) 
and the lifetime of the state to set this cut-off (see PS64).
P1 discussed the possible cut-offs that can be employed, and compared H~I emission spectra
for each. The uncertainties imposed by the cut-offs are in addition to the differences with the SC 
approaches in the cross sections. 
Overall, we showed that the rate coefficients obtained from QM vs SC theories 
result in differences of up to 10\%  for H~I recombination line spectra. We recommended that 
PS64 be used for dipole transitions because of its speed of evaluation and good agreement
with the QM results of \citet{Vrinceanu2001}. 

This paper re-examines the theory of $l$-changing 
collisions for the more complex and important case of He~I Rydberg states. 
In the case of H-like systems we deal with only one electron and all $l$ 
subshells can be taken to be energy degenerate.
The He~I case presents a richer phenomenology with its two spin systems  where large-$l$ 
subshells are nearly degenerate, while small-$l$ subshells are non-degenerate. 
Different approaches are needed to treat collisions for each of these cases.
The degenerate and non-degenerate treatments are reviewed in Section  \ref{sec:lchanginghe}. 
In Section \ref{sec:PS-M}, we introduce a modification to the PS64 approach 
at small impact parameters, guided by the VOS12-QM probabilities, which gives improved
accuracy in circumstances sensitive to close collisions, i.e.\@ low temperatures and/or high densities.
In Section \ref{sec:sims}, we test the influence 
of our results on the final spectrum by calculating He~I  recombination 
emissivities for different 
densities using Cloudy, the spectral simulation code last described by  
\citet{CloudyReview13}. We  recommend the best rates to be used in simulations 
and abundance determinations, and present 
comparisons of the He~I lines obtained using different data, in Section \ref{sec:results}.

\section{Atomic physics of \MakeLowercase{$l$}-changing collisions}
\label{sec:lchanginghe}
We focus on $l$-changing collisions within the same $n$-shell. 
These can be caused by electron collisions or by
the angular momentum Stark-mixing created by 
a slow-moving heavy charged-particle such as a proton or helium nucleus. 
Electron collisions (energy changing) dominate transitions between non-degenerate $l$-subshells 
\citep{Brocklehurst1972}.
Heavy particles will dominate when there is little or no energy exchange 
and they are effective at small impact velocities and large distances.
This requires the $l$-subshells to be energy degenerate, or nearly so. 

Different approaches must be  adopted for different $l$-subshells in He. 
Unlike  the H case, $LS$-coupling in He gives rise to non-degenerate $l$-subshells. 
Although large-$l$ states may be nearly degenerate in energy, smaller ones ($l<3$) are highly non-degenerate. 
Moreover, energy degenerate $l$-changing 
collision theories overestimate rate coefficients when applied to non-degenerate transitions. 
In this  ($l<3$) case the \cite{Seaton1962} formalism (S62) is used and 
adapted to proton collisions  \citep{Benjamin1999,Porter2009}. 
Larger $l$ are `nearly degenerate' so it is necessary to review the QM
cut-off criteria adopted in P1 for degenerate $l$-subshells.
This is analyzed in Section \ref{sec:cut-off}.

We have considered the competing theories  for energy
degenerate collisions discussed in P1. As with H, their 
predicted collisional rate coefficients for He do not agree with each other. These are shown 
in Fig. \ref{f:degcrit}, where, as in the H case, the original
calculations from PS64 agree with VOS12 QM calculations 
(see Section \ref{sec:crosssections} for details). 
Both are larger than the SC data given by the same authors. 

\subsection{$l$-changing collision approaches}
\label{sec:crosssections}

In $l$-changing collisions between heavy charged-particles and an atom, the angular 
momentum of the electron 
within an $n$-shell is changed by the Stark effect caused by the electric field 
of the distant slow-moving  particle (typically a proton or an alpha particle) 
so that $nl\rightarrow nl^\prime$. These collisions redistribute $l$-subshell
populations after recombination and can affect how electrons cascade to
ground, changing the line emissivities. 
In P1 we showed that the use of different $l$-changing data led to
differences in the final emissivities in H~I of up to 10\%, especially for transitions
from high $n$ to low $n$ . 
    
In this paper we use the following notation for the various theories.
The impact parameter Bethe-Born approximation of \cite{PengellySeaton1964} 
is called PS64. New QM and SC 
treatments were proposed by \cite{Vrinceanu2001,Vrinceanu2001b}. Following the 
nomenclature from P1, we refer to these theories as VOS12-QM and VOS12-SC 
respectively\footnote{This notation comes about because these formalisms are 
summarized in \cite{VOS2012} who also give computationally convenient expressions.}. The 
latter was further simplified in \cite{VOS2012} by approximating the scaled 
angular momentum, $l/n$, as a continuous variable in order to provide a simple 
analytic equation for rate coefficients.
We denote this method VOS12-SSC (where SSC stands for Simplified Semi-Classical). 
These theories produce very different results.  SC rate coefficients 
are smaller than VOS12-QM and PS64 by a factor of $\sim 6$ while the last two are in good agreement. 

Two issues affect our choice of methods. First, the probabilities 
predicted by QM methods are divergent for dipole ($\Delta l=1$) transitions at large impact
 parameter. A cut-off based on physical collective effects was chosen to truncate
the probability integral. The SC treatment does not have this problem but it fails to describe long-range 
probabilities which QM methods do describe, making them smaller than the QM 
cross sections. 
We note that all of our H- and He-like recombination predictions
in the series of papers culminating in 
\citet{Porter.R12Improved-He-I-emissivities-in-the-case-B-approximation} 
were based on VOS12-SC from \citet{Vrinceanu2001}, while previous work by the UCL group,
culminating in \citet{Storey1995}, used PS64.

\subsection{Cut-offs for non-degenerate levels at large impact parameters}
\label{sec:cut-off}

Two cut-offs were proposed by PS64 and discussed in P1. The first is:
\begin{equation}
R_\text{c}^d  = 0.72 v\tau\,,
\label{eq:Rc1}
\end{equation}
%
% Leaving a blank line leads to TeX thinking there is a new paragraph,
% \noindent avoids the indent, but bad vertical placement remains =>
% use comment lines starting with '%' if spacing is desired.
%
with $v$ being the projectile speed and $\tau$ the lifetime of the initial state.
This is motivated by that fact that the initial state must not radiate before the 
collision is complete. The second possible cut-off is the Debye screening length,
\begin{equation}
R_\text{D} = \left[\frac{\kb T_e}{4\pi e^2n_e}\right]^{1/2}\,,
\label{eq:RD}
\end{equation}
where $e$ is the electron charge, $\kb$ is the Boltzmann constant, $T_e$ is the 
electron temperature  and $n_e$ is the electron density. 
The smallest of these cut-offs sets the maximum impact parameter for which
the collision can occur. 

\begin{figure*}
\begin{center}
\includegraphics[width=0.85\textwidth,clip]{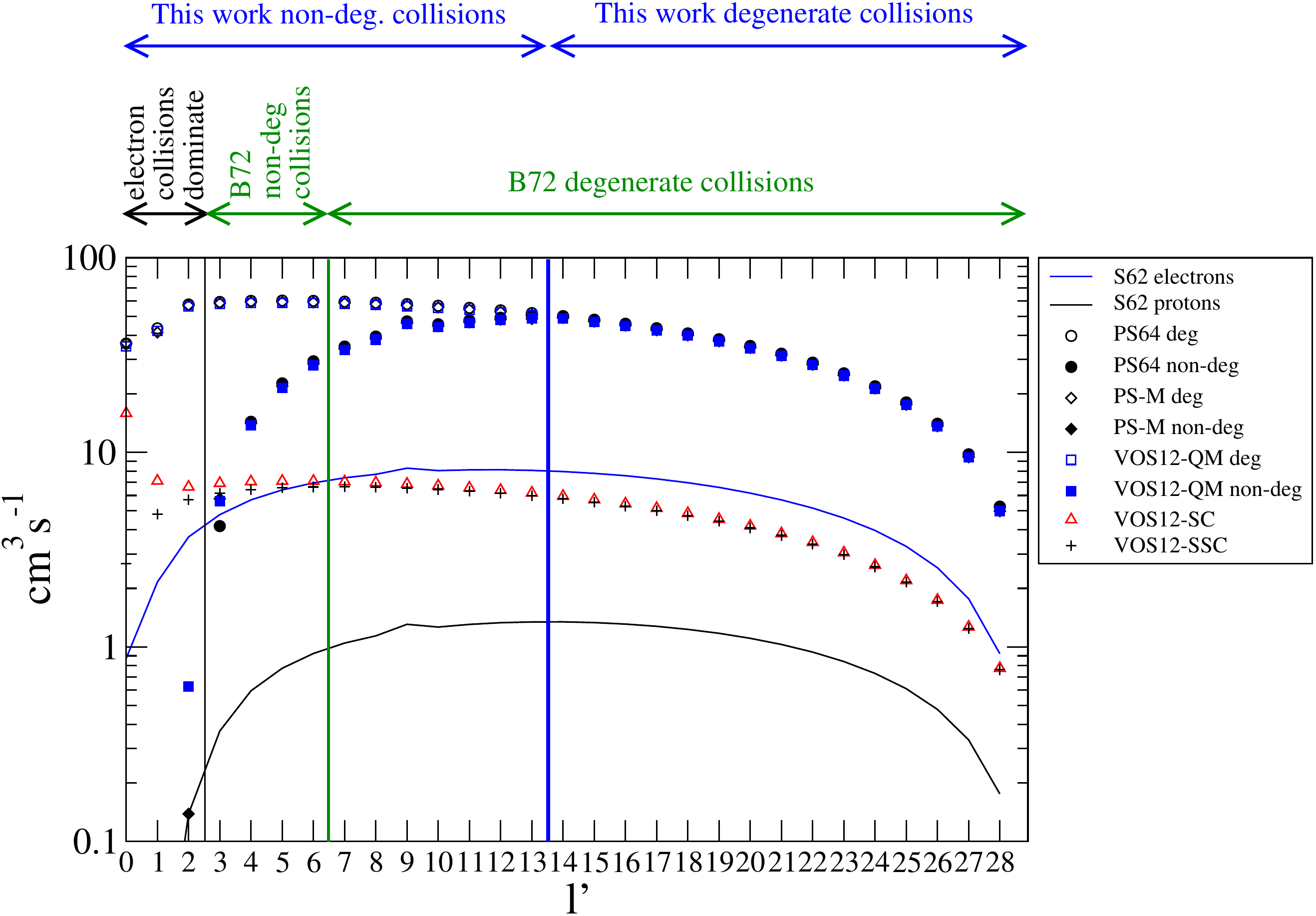}
\caption{\label{f:degcrit} 
Comparison of $l$-mixing collisional rate coefficients for 
H$^+$+He$(n=30)$ singlet collisions for $\Delta l=l-l^\prime=1$ as a function of the final subshell 
$l^\prime$ for $T=10^4\text{K}$ and 
$n_\text{H}=10^4\text{cm}^{-3}$. The B72 criterion from \eqref{eq:B72crit} is satisfied 
usually for $l > 7$ ($l^\prime > 6$) (see the text). In the graph are shown also the limits for electron 
collisions dominating over protons, and the point where cut-offs from equations
\eqref{eq:Rc1} and \eqref{eq:Rc2} are equal. Blue solid line: S62 electron 
collisions; black solid line: S62 proton collisions; open black circles: original 
PS64 degenerate levels calculations; full black circles: original PS64 
calculations using non-degenerate cut-off; open black diamonds: PS-M degenerate levels calculations; 
full black diamonds: PS-M calculations using non-degenerate cut-off;
 open blue squares: VOS12-QM degenerate levels calculations; full blue squares: 
VOS12-QM calculations using non-degenerate cut-off; open red triangles: VOS12-SC;
black + symbols: VOS12-SCC.
}
\end{center}
\end{figure*}

If the energy difference between the initial and final states is large 
compared with the line-widths, $\hbar/\tau$, PS64 replace equation \eqref{eq:Rc1} by an energy 
dependent cut-off:
\begin{equation}
R_\text{c}^{nd} = 1.12\hbar v/\Delta E \,.
\label{eq:Rc2}
\end{equation}
The question of where the 
non-degeneracy is large enough to apply equation \eqref{eq:Rc2} is summarized in Fig.%
\ref{f:degcrit}. For $l \geq 3$ \citet{Brocklehurst1972} 
proposed  using the following inequality to select those $l$-changing collisions where PS64 
cut-offs for degenerate transitions of equations \eqref{eq:Rc1} and \eqref{eq:RD} can be applied:
\begin{equation}
\beta=\frac{L\Delta E}{\hbar 2 W}<0.4\,.
\label{eq:B72crit}
\end{equation}
In equation \eqref{eq:B72crit}, $L$ and $W$ are the mean angular 
momentum and energy of the projectile. This implies 
that the collision time should be less than forty per cent of the spontaneous 
transition time. This is usually fulfilled for $l >7$. 

Fig. \ref{f:degcrit} compares different $l$-changing collision rate coefficients 
for H$^++$He ($n=30;l\to l^\prime=l-1$). The vertical lines indicate the $l'$-ranges where 
the different collision theories apply. The first range 
($l'=0-2$) is where $l$-changing collisions by electrons dominate. 
The next range ($l'=3-6$) covers those $l'$ where equation \eqref{eq:B72crit}
is not satisfied and the $l$-subshells are non-degenerate, but 
$l$-changing proton collisions dominate over electron ones. In the final range ($l'=7-28$) 
the inequality \eqref{eq:B72crit} is satisfied. Finally, the thick blue vertical bar 
beyond $l'=13$ shows the $l^\prime$ 
from where the PS64 rates, using both the degeneracy cut-off from eq
\eqref{eq:Rc2}
and the non-degenerate cut-off from equation \eqref{eq:Rc1} give the same result.
For lower $l$, $R_\text{c}^{nd}<R_\text{c}^{d}$ and for higher $l$, 
$R_\text{c}^d\leq R_\text{c}^{nd}$. Using the non-degenerate cut-off for $l'\leq13$
and the degenerate one for $l'>13$ 
ensures a smooth merging of the 
results. For this reason, we  decided to use the latter as a criterion to separate 
degenerate and nearly degenerate subshells. 
In our calculations we use the QM methods with the minimum
 of the cut-offs from equations \eqref{eq:Rc1}, \eqref{eq:RD} and \eqref{eq:Rc2}
when $l\geq3$:
\begin{equation}
R_\text{c} = \min\{R_\text{c}^d,R_\text{c}^{nd},R_\text{D}\}\,.
\label{eq:Rcfinal}
\end{equation}
If $l<3$, the energy splitting is large enough for electron collisions to dominate. Electron-impact 
collisions have been calculated using the S62 impact parameter method (blue line in Fig. 
\ref{f:degcrit}) and are clearly larger than the non-degenerate QM proton impact rate coefficients. For 
$l\geq 3$, $l$ subshells are much closer in energy and $\Delta E\to 0$ as $l\to\infty$. Note that the
SC results are comparable to electron collisional ones even at this high $n$ shell ($n=30$), as shown in Fig. 
\ref{f:degcrit}. For the non-degenerate $l$ subshells  ($l<3$), we use the adapted S62 method to 
derive proton collisional rate coefficients (solid black line in Fig. \ref{f:degcrit}). This 
guarantees that the $l$ -changing collisional rate is a smooth function of $l$ where collision are 
primarily driven by electrons for $l<3$ and then by protons for high $l$ (see Fig. \ref{f:degcrit}).

Our criterion differs from the one employed by 
\citet{Benjamin1999}, where the S62 method was used for low $l$ until it was within
 6\% of the PS64 degenerate $l$-changing collisional rates. 
 We found that, at high $n$,
this never occurs, as seen in Fig. \ref{f:degcrit}.

\subsection{Modified Pengelly and Seaton (PS-M)}
\label{sec:PS-M}

\cite{PengellySeaton1964} used the impact parameter method to describe the cross section, 
$\sigma_{ji}$, for the transition $i\rightarrow j$

\begin{equation}
\sigma_{ji} = 2\pi \int^\infty_0 P_{ji}(R) R \text{d} R\,,
\end{equation}
in terms of the probability $P_{ji}$ and impact parameter $R$.

On restricting to dipole transitions ($l\rightarrow l'=l\pm 1$), they made use of the 
Bethe form of the Born approximation for $P$:

\begin{equation}
P_{ji}(R)=\frac{4Z_p^2}{3v^2\omega_i}\frac{S_{ji}}{R^2}\,,
\label{PSline}
\end{equation}
where $S_{ji}$ is the atomic line strength, $\omega_i$ the statistical weight of the initial state $i$, 
and $Z_p$ and $v$ are  the projectile charge and velocity, respectively.
Omitting the spin statistical weight from $S_{ji}$, then $\omega_i\rightarrow \omega_l=(2l+1)$.

This approach gives rise to divergent cross sections as $R\rightarrow 0$ and $R\rightarrow \infty$.
Treatments to cut-off the probability at large impact parameters ($R_\text{c}$) were discussed in
Section \ref{sec:cut-off}.
PS64 introduced a cut-off, $R_1$, at small impact parameters defined by requiring continuity with 
an upper bound to the transition probability of
\begin{equation}
P=\frac{1}{2} \quad\quad \mbox{for} \quad\quad R\le R_1\,,
\end{equation}
where they assumed that in reality the probability for $R<R_1$ followed a rapidly oscillating 
Born-like $\sim\sin^2$ behaviour, but that it was sufficiently accurate to represent the mean behaviour 
when determining the overall collision cross section.
Initially, PS64 took $P=P_{ji}$ but in deriving their formula (43) they took $P=\sum_{j\ne i} P_{ji}$.

\begin{figure}
\begin{center}
\includegraphics[width=0.4\textwidth,clip]{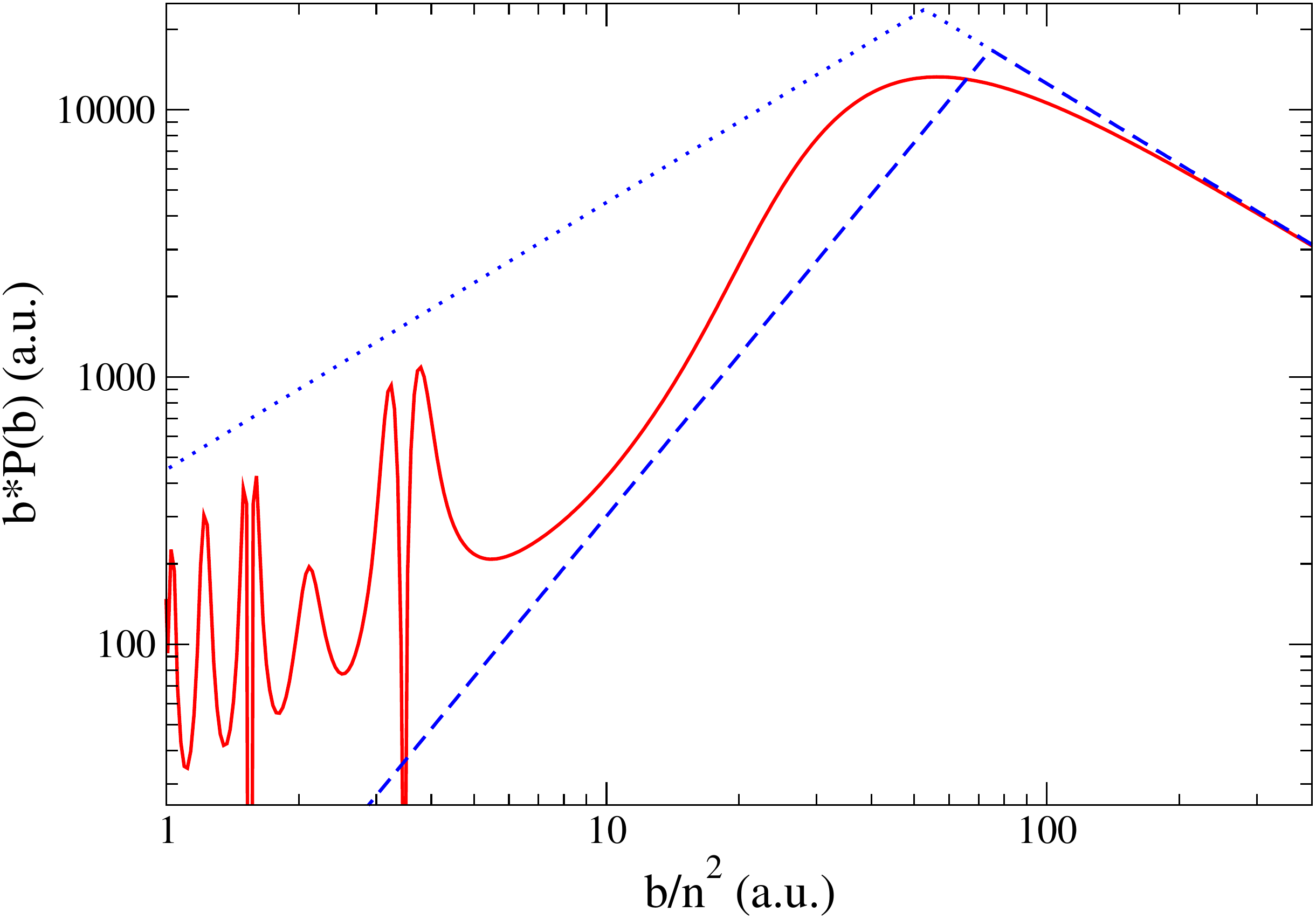}
\caption{\label{f:bpb_imp} Probability times the impact parameter versus scaled impact parameter 
for H$^+$+He($n=30$; $l=29\to28$), at $v=0.25/n$. Solid line, VOS12-QM; dotted line, PS64 ($P_1=1/2$ ) ; 
dashed line, modified PS-M probabilities from equation \eqref{PPSM}, and \eqref{PSline}.}
\end{center}
\end{figure}

The detailed QM results of  \cite{Vrinceanu2001}  --- see Fig.\@ \ref{f:bpb_imp} --- do indeed show 
that the transition probability saturates at some $R$, and undergoes oscillations as $R$ reduces further. 
However, for $n>1$ the envelope of the probability oscillations decreases as $R$ becomes smaller, 
rather than remaining constant as assumed by PS64.  This suggests
the following modification for $R\le R_1 \le R_\text{c}$:
\begin{equation}
P(R)=P_1 \frac{R}{R_1}\,,
\label{PPSM}
\end{equation}
where 
\begin{equation}
2P_1R^2_1=\frac{8Z_p^2\mu I_\text{H}}{3E\omega_l}S_{ji}\,,
\label{Ronem}
\end{equation}
and $\mu$ is the reduced mass, $E$ the energy in the centre-of-mass reference frame, and $I_\text{H}$ 
is the Rydberg energy. Fig.\@ \ref{f:bpb_imp} shows that
this choice of probability better matches the QM probability in the range  $b/n^2=10-100$, 
and to which low temperature rate coefficients are sensitive.

Now, define $E_\text{min}$ to be the energy at which $R_1=R_\text{c}$. Then, for $E \ge E_\text{min}$, we have 
\begin{equation}
\sigma_{ji} =  \frac{\pi a^2_0 \mu I_\text{H}}{2\omega_l E} D_{ji}
\left[\frac{2}{3}+2\log\left(\frac{R_\text{c}}{R_1}\right)\right]\,.
\label{QPSM2}
\end{equation}
Here $a_0$ is the Bohr radius. The factor $\frac{2}{3}$ in (\ref{QPSM2}) arises from the linear form of (\ref{PPSM})
and taking $P_1=\frac{1}{2}$ and summing-over $j$ in (\ref{Ronem}) regains the original PS64 formula (for a constant $P$).
Note, PS64 consider an unresolved $l\to l\pm 1$ cross section, i.e. summed-over final $l'$.
We discuss historic modifications of the original PS64 formula to obtain a final-state resolved form,
and the lack of reciprocity that the use of an $R_1$ engenders, in Appendix \ref{ADNL}.

Writing
\begin{equation}
D_{ji}=\frac{8Z_p^2}{3a_0^2}S_{ji},
\end{equation}
we have
\begin{equation} 
D_{nl\to nl^\prime}=\left(\frac{Z_p}{Z_t}\right)^26n^2l_>\left(n^2-l_>^2\right)\,,
\label{eq:Dnl}
\end{equation}
where $Z_t$ is the target charge and $l_> = \text{max}(l,l^\prime)$ and $l^\prime=l\pm 1$ still. 

The rate coefficient $q_{ji}$ is obtained by convolving the cross section with a Maxwellian
energy distribution. The lifetime or energy splitting cut-off ($R_\text{c}=\min(R^d_\text{c},R^{nd}_\text{c})$) depends on the scattering
energy ($R_\text{c}\propto \sqrt{E}$) while the Debye limit ($R_\text{c}=R_\text{D}$) is independent of it. PS64 assumed a mean cut-off by using 
one or the other at all energies, based-on the smallest at $3v_{\text{RMS}}/4$ for example.
Quadrature of the QM cross sections simply switches between the two forms as appropriate.
We can implement the same approach analytically by splitting the convolution integral into two parts:
define $E_\text{c}$ be the energy such that $R_\text{D}=\min(R^d_\text{c},R^{nd}_\text{c})$,
then 

\begin{equation}
q_{ji}= \frac{a^3_0}{\tau_0\omega_l}\left(\frac{\pi \mu I_\text{H}}{\kb T}\right)^\frac{1}{2} D_{ji}
\left[ \frac{2}{3}\rme^{-\Ubar_\text{m}} + 2E_1(\Ubar_\text{m}) - E_1\left(U_\text{c}\right)
\right]\,,
\label{QPMS5}
\end{equation}
where $\Ubar^2_\text{m}=U_\text{m} U_\text{c}$, $U_\text{m}=ER_1^2/(R_\text{c}^2 \kb T)=E_\text{min}(\mbox{Debye})/ \kb T$, $U_\text{c}=E_\text{c}/\kb T$, $E_1$ is the first
exponential integral and $\tau_0=\hbar/I_H = 2.4188\times10^{-17}$~s is the atomic unit of time.
Note, taking $\Ubar_\text{m}=U_\text{m}=U_\text{c}$ corresponds to using the Debye cut-off at all energies while
$E_1(U_\text{c}\rightarrow \infty) \rightarrow 0$ corresponds to using the lifetime/splitting one.
PS64 made the additional assumption $E_\text{min}\ll \kb T$ and expanded the exponential factors to
leading order. We do not.

This rate coefficient (\ref{QPMS5}) neglects the contribution from cross sections with 
$E < E_\text{min}$, i.e. $R_\text{c} < R_1$ by definition. This was reasonable for PS64 to do since $P=1/2$ there.
Now that we have a more reasonable representation of $P(R<R_1)$ it makes sense to include it.
Then, the cross sections arise only from integration over probabilities given by \eqref{PPSM}, with $R_1$ 
replaced\footnote{Use of $R_1$ here leads to
$P(R)\rightarrow 0$ as $E\rightarrow 0$ (since
$R_1 \rightarrow \infty$ then) which is not born out by the QM $P(R)$.}
by $R_\text{c}$.
Then,
\begin{equation}
\sigma_{ji} (E\le E_\text{min}) =  \frac{2}{3}\pi P_1 R^2_\text{c} \,.
\label{QPSE1}
\end{equation}
This is readily convolved with a Maxwellian distribution to give the {\it additional}
contribution to the rate coefficient, given by (\ref{QPMS5}), from $0\le E\le E_\text{min}$:
\begin{align}
q_{ji}(E\le E_\text{min})=& \frac{a^3_0}{\tau_0\omega_l}\left(\frac{\pi M I_\text{H}}{\kb T}\right)^\frac{1}{2} D_{ji}\nonumber\\
&\times \frac{2}{3}\left[2-2\rme^{-\Ubar_\text{m}}(1+\Ubar_\text{m}+\frac{1}{2}\Ubar^2_\text{m})\right]/\Ubar_\text{m}^2\,.
\label{QPSM6}
\end{align}
We note that the cut-off assumed here is the energy dependent lifetime/splitting one.
In the highly unlikely case that 
\begin{equation}
U_\text{c}<\Ubar_\text{m} \quad\quad \mbox{and, hence,}\quad\quad U_\text{c}<U_\text{m}\,,
\label{eq:cond}
\end{equation}
then the convolution integral is easily split into two again. However, a search of the parameter space 
has not given a case where the condition \eqref{eq:cond} was satisfied.
The contribution from equation \eqref{QPSM6} varies, being
 typically $\sim5\%$ of \eqref{QPMS5}, and thus is important for accurate calculations.
At very low 
temperatures and very high densities, $T\lesssim 100$~K and $n_\text{H} \gtrsim 10^{10}\text{cm}^{-3}$ 
(these values depend on the energy difference of the transition and the lifetime of the initial state), 
the contribution
from equation \eqref{QPSM6} might be larger than equation  \eqref{QPMS5}, which is the `main' term.
In that case, more accurate approaches may be necessary to be taken in 
account for the calculation of the probabilities.  

Following a widespread comparison of PS-M and QM results on H, we take $P_1=1/4$, corresponding to a constant
`branching ratio' of $1/2$ in that picture --- see Appendix \ref{ADNL}. But, use of an $l$-dependent
branching ratio gives only marginally worse results overall at low-$l$. 
Note, this validates PS64's final choice of $P=\sum_{j\ne i} P_{ji}=1/2$ rather than $P=P_{ji}$.
A comparison between selected $l$-changing rate coefficients is given in Table \ref{t:rates}. 

The SC rate coefficients do not have imposed cut-offs and do not depend on density.
They depend linearly on temperature.
The Debye cut-off makes the QM rate coefficients decrease
as the density increases (as seen at high $l$), and they have a more complicated temperature dependence
due to structure within the probability distribution. 
The ratio between the PS-M rate coefficients and the `standard' PS64 ones
increases for higher densities and lower temperatures but this difference is only a few 
percent, except for low $l$ and low temperatures, where the energy separation of the $l$ levels makes the large cut-off
$R_\text{c}$ comparable to the small cut-off $R_1$. As expected, the modification of the small impact parameter probability
causes PS-M to be in much better agreement with VOS12-QM rate coefficients for all 
temperatures and densities, although there are still large differences at $T=10^2K$ and $l=4\to l^\prime=3$ where 
the PS-M rate coefficients overestimate the VOS12-QM ones by a factor $\sim10$. These differences, shown in Table \ref{t:rates},
could produce a significant effect on line intensities as seen in Fig. \ref{f:linesratio}. 

\begin{table*}\footnotesize
\begin{center}
\caption{\label{t:rates} 
Comparison of rate coefficients ($\text{cm}^{3}\text{s}^{-1}$) from
 the different theoretical SC and QM methods for $n=30$ and low- and high-$l$ for
 different temperatures and densities. PS64: `standard' PS64 from equation 
\eqref{eq:PS64}; PS-M: equation \eqref{QPMS5}; VOS12-QM: QM method from
 equation (2) in \protect\cite{VOS2012}; VOS12-SC: SC method from equation (6) in \protect
\cite{VOS2012}; VOS12-SSC: Simplified SC method from equation (9) in
\protect\cite{VOS2012}. The relatively large energy separation of the $l=4\to l^\prime=3$ transition leads 
to $R_\text{c} < R_1$ at $T=100$K  and the `standard' PS64 formula gives a negative result.}
\begin{tabular}{c c c c c c c c c c c |}
\cline{3-11}
 & & \multicolumn{3}{|c|}{$n_\text{H}=10^2\text{cm}^{-3}$}
&\multicolumn{3}{|c|}{$n_\text{H}=10^4\text{cm}^{-3}$}
&\multicolumn{3}{|c|}{$n_\text{H}=10^6\text{cm}^{-3}$}\\
\cline{3-11}
& & \multicolumn{1}{|c|}{$T_H=10^2\text{K}$}&\multicolumn{1}{|c|}{$T_H=10^4\text{K}$}
&\multicolumn{1}{|c|}{$T_H=10^6\text{K}$}&\multicolumn{1}{|c|}{$T_H=10^2\text{K}$}
&\multicolumn{1}{|c|}{$T_H=10^4\text{K}$}
&\multicolumn{1}{|c|}{$T_H=10^6\text{K}$}&\multicolumn{1}{|c|}{$T_H=10^2\text{K}$}
&\multicolumn{1}{|c|}{$T_H=10^4\text{K}$}&\multicolumn{1}{|c|}{$T_H=10^6\text{K}$}\\
\hline
\multicolumn{1}{|c}{\multirow{5}{*}{$l=4 \to l^\prime=3$}} 
& \multicolumn{1}{|c|}{VOS12-QM} & 1.66e-3 & 5.61e+0 & 3.51e+0 & 1.66e-3 & 5.61e+0 
& 3.51e+0 & 1.66e-3 & 5.61e+0 & 3.51e+0 \\
\cline{2-2}
\multicolumn{1}{|c}{} & \multicolumn{1}{|c|}{PS64} & --- & 4.18e+0 & 3.65e+0 
& --- & 4.18e+0 & 3.65e+0 & --- & 4.18e+0 & 3.65e+0 \\
\cline{2-2}
\multicolumn{1}{|c}{} & \multicolumn{1}{|c|}{PS-M} & 2.00e-2 & 5.77e+0 & 3.57e+0 
& 2.00e-2 & 5.77e+0 & 3.57e+0 & 2.00e-2 & 5.77e+0 & 3.57e+0 \\
\cline{2-2}
\multicolumn{1}{|c}{} & \multicolumn{1}{|c|}{VOS12-SC} & 6.94e+1 & 6.94e+0 & 6.94e-1 
& 6.94e+1 & 6.94e+0 & 6.94e-1 & 6.94e+1 & 6.94e+0 & 6.94e-1 \\
\cline{2-2}
\multicolumn{1}{|c}{} & \multicolumn{1}{|c|}{VOS12-SSC} & 6.17e+1 & 6.17e+0 & 6.17e-1 
& 6.17e+1 & 6.17e+0 & 6.17e-1 & 6.17e+1 & 6.17e+0 & 6.17e-1 \\
\hline
\multicolumn{1}{|c}{\multirow{4}{*}{$l=29\to l^\prime=28$}} 
& \multicolumn{1}{|c|}{VOS12-QM} 
& 3.80e+1 & 6.18e+0 & 8.55e-1 & 2.61e+1 & 4.99e+0 & 7.35e-1 & 1.42e+1 & 3.80e+0 & 6.16e-1 \\
\cline{2-2}
\multicolumn{1}{|c}{} & \multicolumn{1}{|c|}{PS64} & 4.06e+1 & 6.44e+0 & 8.81e-1 
& 2.87e+1 & 5.25e+0 & 7.61e-1 & 1.68e+1 & 4.06e+0 & 6.42e-1 \\
\cline{2-2}
\multicolumn{1}{|c}{} & \multicolumn{1}{|c|}{PS-M} & 3.80e+1 & 6.18e+0 & 8.54e-1 
& 2.60e+1 & 4.99e+0 & 7.35e-1 & 1.41e+1 & 3.79e+0 & 6.16e-1 \\
\cline{2-2}
\multicolumn{1}{|c}{} & \multicolumn{1}{|c|}{VOS12-SC} & 7.76e+0 & 7.76e-1 & 7.76e-2 
& 7.76e+0 & 7.76e-1 & 7.76e-2 & 7.76e+0 & 7.76e-1 & 7.76e-2 \\
\cline{2-2}
\multicolumn{1}{|c}{} & \multicolumn{1}{|c|}{VOS12-SSC} & 7.63e+0 & 7.63e-1 & 7.63e-2 
& 7.63e+0 & 7.63e-1 & 7.63e-2 & 7.63e+0 & 7.63e-1 & 7.63e-2 \\
\hline
\end{tabular}
\end{center}
\end{table*}

\section{Cloudy simulations}
\label{sec:sims}

The impact of these changes in the $l$ subshell 
collisional data on He~I recombination line emission was tested in Cloudy using a model very 
similar to the one used in P1 for H I spectra. Line emissivities 
applying both the QM and SC formalisms are compared to estimate the uncertainty 
created by  differences in the theories considered here. 

\subsection{Description of the model for He}

We use the development version of Cloudy, 
the spectral simulation code last described by
\citet{CloudyReview13}.
The results presented in this paper come from the branch Hlike\_HS87 at revision r11113.

As in the H-only case (P1), a single layer of gas
 has been considered and 
emissivities from recombination lines calculated. A cosmic helium
abundance, He/H =0.1, was  assumed.
The cloud is radiated by a
narrow band of radiation, a `laser', centered at 2 Ryd, with an ionization
parameter of U = 0.1 \citep{AGN3}. The state of the emitting gas in these conditions resembles that in H~II regions, 
observationally relevant for 
the determination of He abundances \citep{Izotov2014}.
In this calculation we set the code to ignore internal excitations that
might have been produced by the incident radiation field.
 A constant gas kinetic temperature of 1\e{4}~K is
 assumed. As in the previous paper, we assume `Case B' \citep{Baker1938}, where
Lyman lines with upper shell $n>2$ are assumed to scatter often enough to be degraded into Balmer lines and Ly$\alpha$. 
The hydrogen 
density is varied over a wide range and the electron density is calculated self-consistently.
The latter is approximately 10\% greater than the hydrogen density since He is singly ionized. 
The atomic data used for He and H emission, except for $l$-changing collisions, 
are the `standard' set of data that has been described in previous works \citep{Porter2005}.

The hydrogen density ranges from $10\ \text{cm}^{-3}$ to $10^{10}\, \text{cm}^{-3}$, 
a range of densities large enough to cover diverse astrophysical environments from the interstellar 
medium to quasars. As in the case of hydrogen, $n$-shell states can be treated 
as `resolved', where all $l$ subshells are modeled separately, or `collapsed'
where the $l$ subshell  population is statistically distributed. 
The chosen maximum principal quantum number for resolved levels, 
$n_r$, is density dependent (PS64), 
with $n_r\sim 60$ at the lowest density and $n_r=20$ at the highest 
density.

A set of 411 of the brightest He~I lines has been selected in a range from 2000\AA\ to 30$\mu$m 
covering most of the spectral range of today's observations. These are both 
singlet and triplet transitions, where the upper levels have principal quantum numbers 
between $n=3$ and $n=36$, and the lower levels have $n$ between $2$ and $8$. 

\subsection{Results}
\label{sec:results}

\begin{figure*}
\begin{center}
\includegraphics[width=0.5\textwidth,clip]{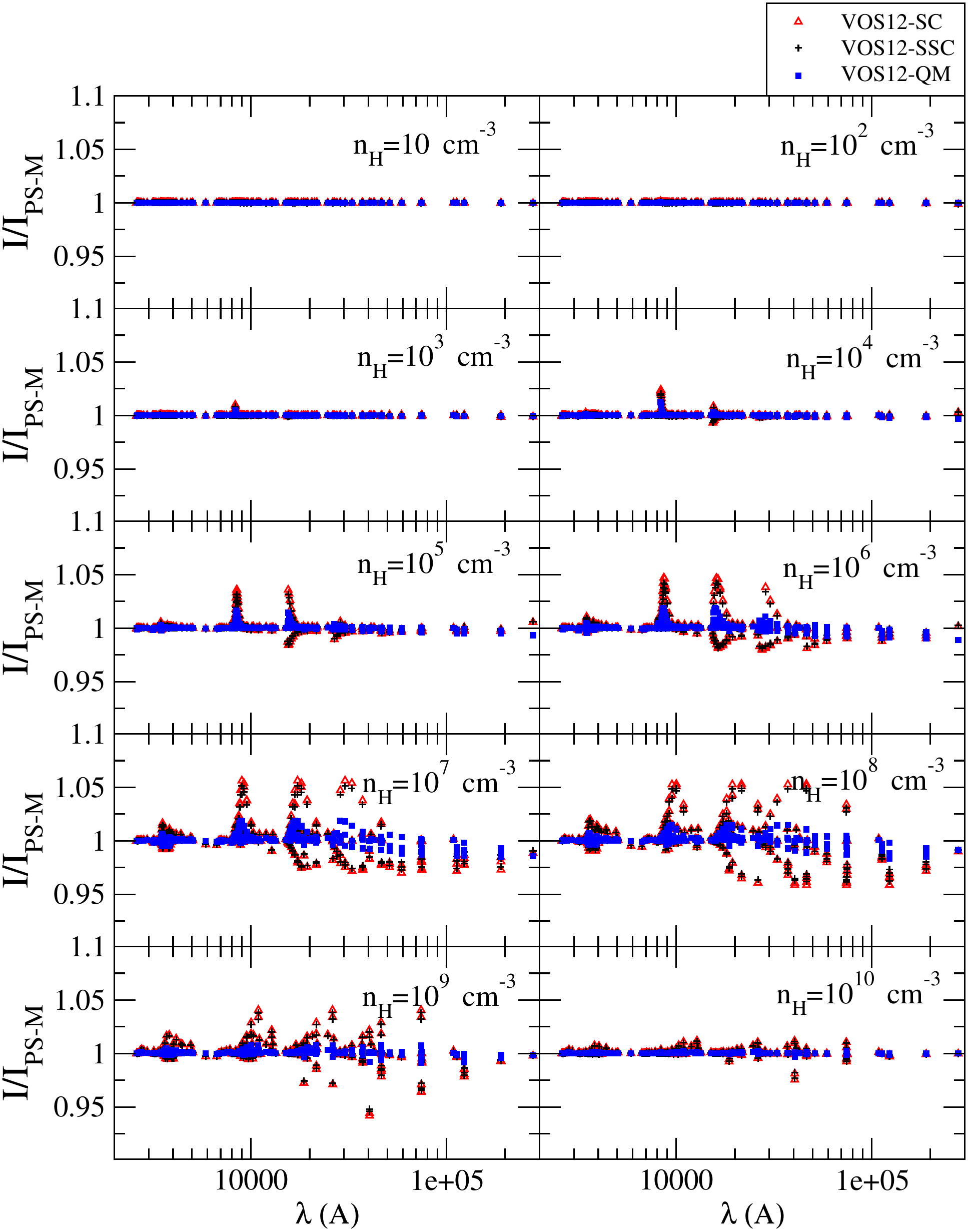}
\caption{\label{f:linesratio} Ratios of He~I lines for Cloudy simulations using 
the different data sets considered in this work with respect to the PS-M approach,
for $T=10^4$K.}
\end{center}
\end{figure*}

As in P1, a comparison of line emissivities will  show the 
behaviour of different $l$-changing theories. Predicted emissivities, normalized to the
 PS-M theory, are presented in Fig. \ref{f:linesratio}  for each of the 
theories described in P1 and discussed here. Density is increasing from upper left to bottom right.

As in the case of H~I, changes in the $l$-changing data have little impact at
very low densities, where radiative decays are much faster than $l$-changing collisions.
The predictions do change between densities of 
$n_\text{H} = 10^4 - 10^{10} \text{cm}^{-3}$, where $l$-mixing processes compete with radiative 
decay. At the highest densities, collisions start to dominate over radiative processes and  
the $l$ subshells come into LTE. 
Differences in the rate coefficients result in deviations up to $\sim7$\% in many lines. 
Paschen ($\lambda\sim9000$~\AA) and Brackett ($\lambda\sim 15000$~\AA) series are 
clearly visible. 
Note that $l<3$ collisions are considered to be dominated by electron-impact 
and so no differences are expected for the Balmer series except as a result of 
cascading from higher $n$. We find the largest effect at densities around 
$n_\text{H}= 10^5 - 10^9\text{cm}^{-3}$. For all transitions, the larger 
the upper level quantum number $n$ is, the more $l$ subshells are 
available for redistribution and differences in the $l$-mixing collisions data 
become more important. As in the case of H~I, the two simulations based on QM 
calculations (PS-M and VOS12-QM) agree much more closely with each 
other for all densities 
considered than with the two based on SC data (VOS12-SC and VOS12-SSC). The latter have emissivities 
that are  different from PS-M, particularly for lines where the upper level has a high 
principal quantum number $n$. 
This effect is more pronounced for densities of $n_\text{H}=10^7\ \text{cm}^{-3}$ and 
$n_\text{H}=10^8\ \text{cm}^{-3}$ 
where at high $n$ levels, collisional effects 
are strong and comparable with radiative decay. Therefore, differences in the
$l$-changing collisional rates have a larger impact at these densities. 
This also happens on a smaller scale at the immediately lower 
densities of $n_\text{H}=10^5\ \text{cm}^{-3}$ and $n_\text{H}=10^6\ \text{cm}^{-3}$, where the smaller 
collisional rates produced by the SC approaches lead to noticeable differences in 
emissivities.

\subsubsection{Impact on observed He~I low $n$ lines}
 
We have also examined the impact of the various $l$-changing theories on the 
emissivities of the most commonly observed lines at visible wavelengths. These lines 
have relatively small upper $n$, typically 3 or 4, 
therefore low-$l$ dipole transitions, with these being dominated by $ns$, $np$, and $nd$ subshells,
so it is expected that 
the various $l$-changing heavy impact collisions will only affect these lines through  
cascades from higher levels which are affected. Table \ref{t:somelines} lists emissivities of some  
important lines  
normalized to PS-M.  These are given as a function of  temperature and density for the 
different cases considered here. For these representative low-lying lines, 
$l$-changing collisions have a greater effect at low temperatures where 
the Stark mixing from the projectile is more effective. 
There are also larger differences at high densities, 
where collisional processes dominate over radiative.
%the QM cut-offs are smaller and low impact parameter differences have a higher role.
The differences between VOS12-QM and PS-M are generally smaller than 1\%. However, 
at low temperatures the 4471.49\AA\ predicted intensities disagree by $\sim$2\%, presumably as a 
consequence of the cumulative effects of the differences in rates seen in Table \ref{t:rates}. 
The SC results can  disagree by up to $\sim$10\% for 4471.49\AA\ line and by up to $\sim$5\% for 
7065.22\AA\, 5875.64\AA\ and 6678.15\AA\ lines, for $T=100K$ and $n_\text{H} = 10^4$. These differences can 
increase by up to $\sim$14\%, $\sim$11\%, $\sim$10\% and $\sim$10\%, respectively, for the 
extreme case of $n_\text{H} = 10^6\text{cm}^{-3}$ and 
$T=100K$.
None of the 
lines in Table \ref{t:somelines} correspond to a transition with $l\geq 3$, 
so $l$-changing transitions are dominated by electron collisions. However, the 
upper levels of the first three lines in table \ref{t:somelines} arise from 
$l=2$ which  receives electrons for $\Delta l=1$ transitions from $l=3$ 
subshells that are partially populated by Stark effect $l$-changing mechanisms. 
Cascade effects from higher $l$ are important for the 
 1s3s $^1$S -- 1s2p $^1$P transition at 7281.35\AA .
The 10830.31\AA\ intrashell triplet-triplet transition 1s2s $^3$S -- 1s2p $^3$P  
is less likely to be affected by differences in $l$-changing 
collisions since the emission line is mainly produced by electron collisions from the metastable
lower level, and cascade effects 
make only a small contribution. 

%%%%%%%%%%%%%%%%%%%%%%%%%%%%%%%%%%%%
\begin{table*}\footnotesize
\begin{center}
\caption{\label{t:somelines} Some representative emissivities for the He~I spectrum, normalized to 
PS-M results, at various densities and temperatures. The VOS12 labels are as in table \ref{t:rates}.}
\begin{tabular}{c c c c c c c c c c c |}
\cline{3-11}
 & & \multicolumn{3}{|c|}{$n_\text{H}=10^2\text{cm}^{-3}$}&\multicolumn{3}{|c|}
{$n_\text{H}=10^4\text{cm}^{-3}$}&\multicolumn{3}{|c|}{$n_\text{H}=10^6\text{cm}^{-3}$}\\
\cline{3-11}
& & \multicolumn{1}{|c|}{$T_H=10^2\text{K}$}&\multicolumn{1}{|c|}{$T_H=10^4\text{K}$}
&\multicolumn{1}{|c|}{$T_H=10^6\text{K}$}&\multicolumn{1}{|c|}{$T_H=10^2\text{K}$}
&\multicolumn{1}{|c|}{$T_H=10^4\text{K}$}&\multicolumn{1}{|c|}{$T_H=10^6\text{K}$}
&\multicolumn{1}{|c|}{$T_H=10^2\text{K}$}&\multicolumn{1}{|c|}{$T_H=10^4\text{K}$}
&\multicolumn{1}{|c|}{$T_H=10^6\text{K}$}\\
\hline
\multicolumn{1}{|c}{\multirow{2}{*}{4471.49~\AA}} & \multicolumn{1}{|c|}{VOS12-QM} 
& 1.007 & 1.000 & 1.000 & 1.020 & 1.000 & 1.000 & 1.010 & 1.001 & 1.001 \\
\cline{2-2}
\multicolumn{1}{|c}{} & \multicolumn{1}{|c|}{VOS12-SC} & 1.025 & 1.000 & 1.000 & 1.105 
& 1.000 & 0.999 & 1.138 & 1.001 & 0.997 \\
\cline{2-2}
\multicolumn{1}{|c}{1s4d $^3$D -- 1s2p $^3$P} & \multicolumn{1}{|c|}{VOS12-SSC} & 1.024 
& 1.000 & 1.000 & 1.103 & 1.000 & 0.999 & 1.134 & 1.001 & 0.997 \\
\hline
\multicolumn{1}{|c}{\multirow{2}{*}{5875.64~\AA}} & \multicolumn{1}{|c|}{VOS12-QM} 
& 1.001 & 1.000 & 1.000 & 1.003 & 1.000 & 1.000 & 0.999 & 1.000 & 1.000 \\
\cline{2-2}
\multicolumn{1}{|c}{} & \multicolumn{1}{|c|}{VOS12-SC} & 0.992 & 1.000 & 1.000 & 0.959 
& 1.000 & 1.000 & 0.900 & 0.998 & 1.000 \\
\cline{2-2}
\multicolumn{1}{|c}{1s3d $^3$D -- 1s2p $^3$P} & \multicolumn{1}{|c|}{VOS12-SSC} & 0.992 
& 1.000 & 1.000 & 0.961 & 1.000 & 1.000 & 0.902 & 0.999 & 1.000 \\
\hline
\multicolumn{1}{|c}{\multirow{2}{*}{6678.15~\AA}} & \multicolumn{1}{|c|}{VOS12-QM} 
& 0.998 & 1.000 & 1.000 & 0.997 & 1.000 & 1.000 & 0.998 & 1.000 & 1.000 \\
\cline{2-2}
\multicolumn{1}{|c}{} & \multicolumn{1}{|c|}{VOS12-SC} & 0.989 & 1.000 & 1.000 & 0.951 
& 1.000 & 1.000 & 0.901 & 0.998 & 1.001 \\
\cline{2-2}
\multicolumn{1}{|c}{1s3d $^1$D -- 1s2p $^1$P} & \multicolumn{1}{|c|}{VOS12-SSC} & 0.990 
& 1.000 & 1.000 & 0.952 & 1.000 & 1.000 & 0.903 & 0.998 & 1.001 \\
\hline
\multicolumn{1}{|c}{\multirow{2}{*}{7065.22~\AA}} & \multicolumn{1}{|c|}{VOS12-QM} 
& 1.000 & 1.000 & 1.000 & 1.001 & 1.000 & 1.000 & 1.003 & 1.000 & 1.000 \\
\cline{2-2}
\multicolumn{1}{|c}{} & \multicolumn{1}{|c|}{VOS12-SC} & 1.009 & 1.000 & 1.000 & 1.055 
& 1.000 & 1.000 & 1.111 & 1.000 & 1.000 \\
\cline{2-2}
\multicolumn{1}{|c}{1s3s $^3$S -- 1s2p $^3$P} & \multicolumn{1}{|c|}{VOS12-SSC} & 1.009 
& 1.000 & 1.000 & 1.053 & 1.000 & 1.000 & 1.109 & 1.000 & 1.000 \\
\hline
\multicolumn{1}{|c}{\multirow{2}{*}{7281.35~\AA}} & \multicolumn{1}{|c|}{VOS12-QM} 
& 0.998 & 1.000 & 1.000 & 0.992 & 1.000 & 1.000 & 0.997 & 1.000 & 1.000 \\
\cline{2-2}
\multicolumn{1}{|c}{} & \multicolumn{1}{|c|}{VOS12-SC} & 1.002 & 1.000 & 1.000 
& 1.012 & 1.000 & 1.000 & 1.059 & 1.000 & 1.000 \\
\cline{2-2}
\multicolumn{1}{|c}{1s3s $^1$S -- 1s2p $^1$P} & \multicolumn{1}{|c|}{VOS12-SSC} 
& 1.002 & 1.000 & 1.000 & 1.011 & 1.000 & 1.000 & 1.057 & 1.000 & 1.000 \\
\hline
\multicolumn{1}{|c}{\multirow{2}{*}{10830.31~\AA}} & \multicolumn{1}{|c|}{VOS12-QM} 
& 1.002 & 1.000 & 1.000 & 1.005 & 1.000 & 1.000 & 1.002 & 1.000 & 1.000 \\
\cline{2-2}
\multicolumn{1}{|c}{} & \multicolumn{1}{|c|}{VOS12-SC} & 1.000 & 1.000 & 1.000 & 1.003 
& 1.000 & 1.000 & 0.992 & 1.000 & 1.000 \\
\cline{2-2}
\multicolumn{1}{|c}{1s2s $^3$S -- 1s2p $^3$P} & \multicolumn{1}{|c|}{VOS12-SSC} & 1.001 
& 1.000 & 1.000 & 1.003 & 1.000 & 1.000 & 0.992 & 1.000 & 1.000 \\
\hline
\end{tabular}
\end{center}
\end{table*}
%%%%%%%%%%%%%%%%%%%%%%%%%%%%%%%%%%%%%%%%%%%%%%%%%%%%%%%%%%%%%%%

\subsubsection{Effect of multipole l-changing}
\label{sec:multipole}

Multipolar $l$-changing collisions strengthen $l$-mixing at high $n$ and can have an effect 
on emissivities. PS64 assumed multipole ($\Delta l > 1$) $l$-changing rates to 
be negligible compared with dipole ones. VOS12 calculate the multipole 
dependence as $\left(\Delta l\right)^{-3}$ for SC calculations. The results for 
the emissivities presented in Fig. \ref{f:linesratio} and 
Table \ref{t:somelines} are obtained considering all possible multipolar 
$l$-changing transitions for VOS12-QM, VOS12-SC and VOS12-SSC results, while 
only dipole transitions are allowed by PS64 and PS-M. We have done that in 
order to get the best prediction that each of the approaches can provide.

\begin{figure}
\begin{center}
\includegraphics[width=0.5\textwidth,clip]{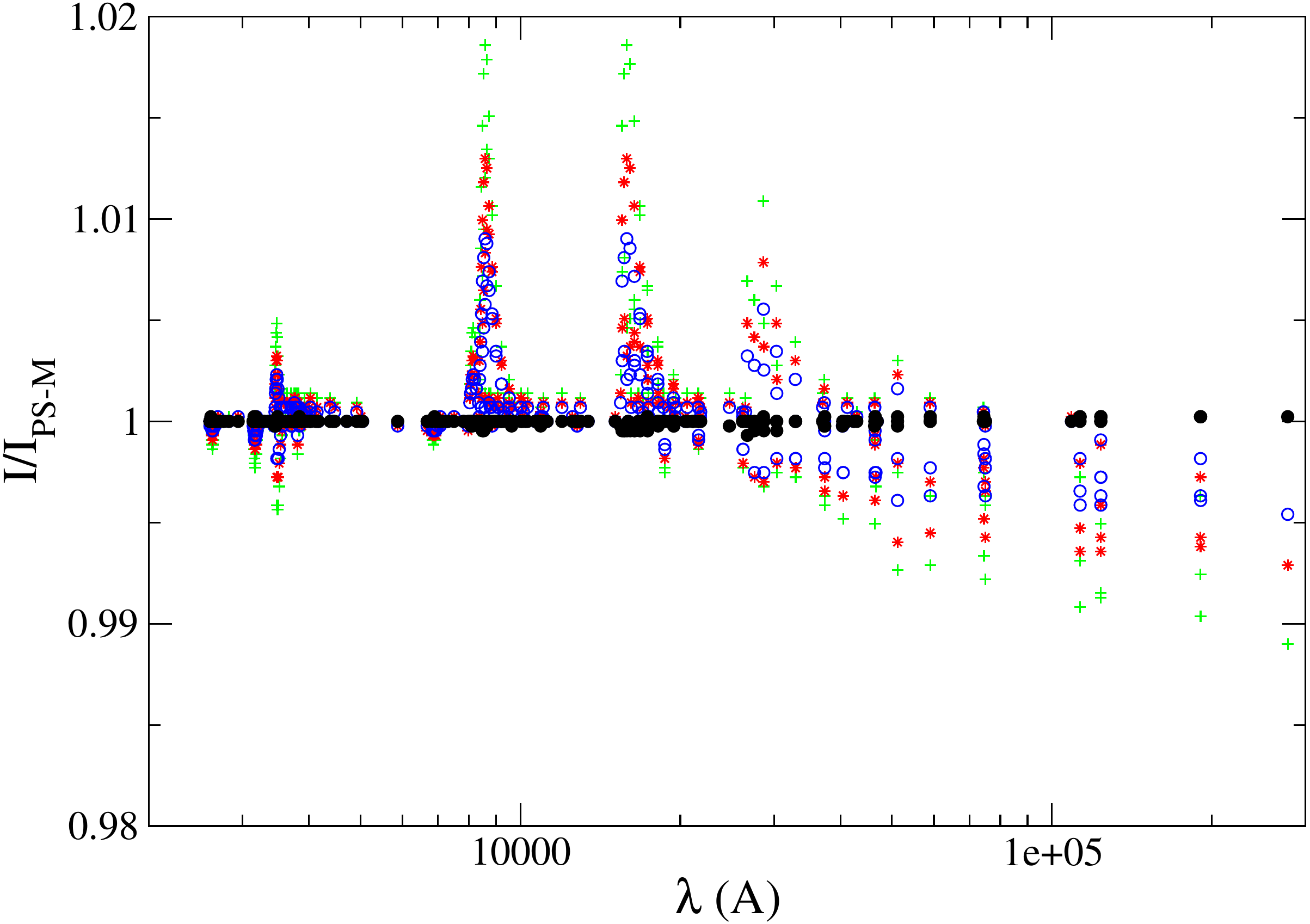}
\caption{\label{f:multipole} Ratios of He~I lines for VOS12-QM to PS-M (dipole) data from
Cloudy simulations including  different multipolar $l$-changing transitions at $T=10^4$K and  
$n_\text{H}=10^6\text{cm}^{-3}$. 
Black filled circles:  $\Delta l = 1$; blue open circles:  $\Delta l < 3$ ; red stars:  $\Delta l < 4$ ; 
green crosses: all multipoles.}
\end{center}
\end{figure}

The effect of multipolar transitions can be quantified by progressively switching-off $\Delta l > 1,2,3$, 
$l$-changing transitions in our calculations at different densities. We have found these 
effects negligible at most densities except for intermediate densities 
($n_\text{H}=10^5-10^6\text{cm}^{-3}$), where differences in emissivities are up 
to $\sim2$\% when these transitions are included in VOS12-QM rates
for $n_\text{H}=10^6\text{cm}^{-3}$. In Fig. \ref{f:multipole}, the multipole effects constitute the 
main contribution to the disagreement between VOS12-QM and PS-M results at 
this density.
Quadrupolar transitions contribute up to $\sim50$\% of the total 
difference of the emissivities with PS-M results. Multipoles have a similar 
contribution to both Paschen and Bracket lines and can also influence Balmer 
lines by cascade effects 
(note that $l\to l^\prime\leq2$ transitions are dominated by electron impact and 
calculated in a similar way in both cases). Similar effects are seen for SC 
calculations.  The overall magnitude of these effects is significantly 
smaller than for the SC/QM differences which are the focus of the present
paper, but are still at a level which would have a significant impact on
the determination of cosmological abundances.  Improving the accuracy of
these predictions in a computationally efficient manner will be the subject
of future work.

\section{Conclusions}

We have applied the different theories of $l$-changing collisions to the 
more complicated case of He. In summary:
%We make the following advances and conclusions:

\begin{itemize}

\item We have improved the accuracy of the PS64 Bethe approximation by modifying the 
probability at small impact parameters so it resembles better the pure QM calculations of 
VOS12. We have not followed the PS64 assumption of $E\gg E_\text{min}$($R_1\ll R_\text{c}$). 
The contribution from small energies to the integration, which was neglected by PS64, 
is taken into account and changes the final rate coefficients by about 5\%.
We call this improved theory PS-M.

\item Cut-offs are important for impact parameter theory.
These highly accurate results can be applied in the non-degenerate case where they are integrated 
to different  large impact parameter cut-offs due the energy differences between intrashell
$l$-subshells. The criteria to decide when to apply degenerate or non-degenerate cut-offs has
been refined to obtain a continuous variation of the rate coefficients as a function of  $l$. 

\item We have compared our improved results with the more accurate 
VOS12-QM results and with the SC theories (which do not account for non-degeneracy within 
$l$ subshells). The new PS-M based emissivities are in very good agreement with the 
pure QM VOS12-QM results and both disagree with SC predictions. 
Cloudy simulations show that the He~I line emissivities predicted by these different
$l$-changing theories change by up to 10\%.
However, emissivity differences of $\sim$2\% remain
between the VOS12-QM theory and the PS-M model at intermediate densities 
$10^4 - 10^8\text{cm}^{-3}$, where radiative and collisional processes compete.
Most of this difference can be attributed to non-dipole $l$-changing collisions which the PS-M approach neglects.    
This should be taken in account when 
testing Big Bang nucleosynthesis models, which require a precision better that 
1\%. 

\end{itemize}

The main result of this work is an improved $l$-changing theory, PS-M, which is highly accurate 
at nearly all densities and temperatures and is computationally fast and easy to implement. 
We recommend using PS-M for $l\geq3$ $l$-changing collisions rates in order to obtain
good accuracy He~I spectral intensities. However, it must be noted that no observational or experimental measurement has been done up to date that supports any of the data sets over the other. Our recommendations are based uniquely on theoretical considerations on the long impact parameter probabilities differences described in P1. Only observations or experiment can definitely set the validity of one of the calculations over the others. 

The PSM approach will be the default in the next release of Cloudy for this type of collisions.
However, we also provide the more precise VOS12-QM method,
which is computationally more expensive,  to allow testing of the various methods and for when 
even higher accuracy results are needed.

\section{Acknowledgments}

We acknowledge support by NSF (1108928, 1109061, and 1412155), 
NASA (10-ATP10-0053, 10-ADAP10-0073, NNX12AH73G, and ATP13-0153), 
and STScI (HST-AR- 13245, GO-12560, HST-GO-12309, GO-13310.002-A, 
HST-AR-13914, and HST-AR-14286.001). MC has been supported by STScI (HST-AR-14286.001-A). 
PvH was funded by the Belgian Science Policy Office under contract no.
BR/154/PI/MOLPLAN.

\bibliographystyle{mn2e}
\bibliography{LocalBibliography,bibliography2.bib}
\bsp

\appendix
\section{Resolved vs unresolved final-states and reciprocity.}
\label{ADNL}
The \cite{PengellySeaton1964} formula  is for an unresolved, by final state, rate coefficient $q_{nl}$
for $nl\rightarrow nl'$ summed-over $l'=l\pm 1$. In general, we require a final-state resolved
expression.

Define $P_1$ of Section \ref{sec:PS-M} by
\begin{equation} 
P_1=\frac{1}{2}B_{ji}
\end{equation}
where the branching ratio $B_{ji}$ is given by 
\begin{equation} 
B_{ji}=\frac{D_{ji}}{w_l D_{nl}}
\end{equation}
and $D_{nl}$, the quantity used by PS64, is given by
\begin{equation}
D_{nl}\equiv\sum_{l'=l\pm 1} \frac{1}{\omega_l}D_{ji}=\frac{Z^2}{z^2}6n^2(n^2-l^2-l-1) \,.
\label{DPS}
\end{equation}
This leads to the final-state resolved form of the PS64 formula
\begin{align}
q_{ji}=&9.93\times10^{-6}\text{cm}^3\text{s}^{-1}\left(\frac{\mu}{m_e}\right)^{1/2} \nonumber \\
\times&\frac{D_{ji}}{\omega_l T^{1/2}}\left[11.54+\log_{10}\left(\frac{Tm_e}{D_{nl}\mu}\right)
+2\log_{10}\left(R_\text{c}\right)\right] \,,
\label{eq:PS64}
\end{align}
as used by Cloudy \citep{CloudyReview13}, \citet{Hummer1987} and \citet{Summers1977}.
The sum over the final $l'$ then recovers the original PS64 unresolved value.
The unresolved PS-M result for arbitrary $P_1$ is obtained simply by summing-over $l'=l\pm 1$.
In general, there is no corresponding closed-expression for it, or need for one.

An issue which affects all forms of the Pengelly and Seaton method is the lack of
reciprocity because of the $D_{nl}$ factor in the $\log_{10}$ term. The standard approach, as used by Cloudy,
is to calculate all rate coefficients in one direction ($l\rightarrow l-1$, say) and obtain 
the rate coefficient for the reverse direction using the principle of detailed balance, viz.
$(2l-1)q_{nl-1\to nl} = (2l+1)q_{nl\to l-1}$ (for $\Delta E_{ll^\prime} \approx 0$).
In general (\citet{Summers1977}), results are not sensitive to the $\log_{10}$ term and,
even if they are, the {\it ad hoc} nature of the original definition of $R_1$ means that the
treatment is unreliable here. 

Comparison of our PS-M $s - p$ rate coefficients calculated in both directions with the QM result
shows that the $p \rightarrow s$  result lies above the QM one while the $s\rightarrow p$ lies
below, by almost the same amount but a little larger. The difference depends strongly on the
value of $R_\text{c}$ compared to $R_1$ but only exceeds $\sim 10\%$ in extreme cases where the plasma
screening is about to delocalise the state. The difference rapidly decreases with increasing $l$.
For helium, we do not use any form of Pengelly and Seaton for $l=0-2$.

\label{lastpage}
\clearpage
\end{document}